\documentclass[conference]{IEEEtran}

\ifCLASSINFOpdf
\else
\fi
\usepackage{graphicx}
\usepackage{tabularx,booktabs}
\usepackage{dingbat}
\usepackage{diagbox}
\usepackage{multirow} 
\usepackage[hyphens]{url}
\usepackage[hidelinks,breaklinks]{hyperref}
\usepackage{xcolor}
\usepackage{amsfonts, amsmath, amsthm, amssymb}
\usepackage{subcaption}
\hypersetup{breaklinks=true}
\usepackage{tikz}
\usepackage{textcomp}
\usepackage{lipsum}
\usepackage{setspace}
\usepackage{verbatim}
\usepackage{svg}

\IEEEoverridecommandlockouts
\newcommand\copyrighttext{%
  \footnotesize \textcopyright 2025 IEEE.  Personal use of this material is permitted.  Permission from IEEE must be obtained for all other uses, in any current or future media, including reprinting/republishing this material for advertising or promotional purposes, creating new collective works, for resale or redistribution to servers or lists, or reuse of any copyrighted component of this work in other works.}
\newcommand\copyrightnotice{%
\begin{tikzpicture}[remember picture,overlay]
\node[anchor=south,yshift=10pt] at (current page.south) {\fbox{\parbox{\dimexpr\textwidth-\fboxsep-\fboxrule\relax}{\copyrighttext}}};
\end{tikzpicture}%
}

\usepackage{graphicx}
\graphicspath{{figures/}}
\begin{document}

\bstctlcite{IEEEexample:BSTcontrol}
%
\title{Evaluations of High Power User Equipment (HPUE) in Urban Environment}

\author{
    \IEEEauthorblockN{Kasidis Arunruangsirilert\IEEEauthorrefmark{1}\IEEEauthorrefmark{3}, Pasapong Wongprasert\IEEEauthorrefmark{2}\IEEEauthorrefmark{3}, Jiro Katto\IEEEauthorrefmark{1}}
    \IEEEauthorblockA{\IEEEauthorrefmark{1}Department of Computer Science and Communications Engineering, Waseda University, Tokyo, Japan}
    \IEEEauthorblockA{\IEEEauthorrefmark{2}Department of Electrical Engineering, Chulalongkorn University, Bangkok, Thailand
    \\\{kasidis, katto\}@katto.comm.waseda.ac.jp, 6670331021@student.chula.ac.th}
    \IEEEauthorblockA{\IEEEauthorrefmark{3}Authors contributed equally to this research.}
    
}
\vspace{-3mm}
%
\maketitle
\vspace{-2mm}
\copyrightnotice
\setstretch{0.97}
\begin{abstract}
While Time Division Duplexing (TDD) 5G New Radio (NR) networks offers higher downlink throughput due to the utilization of the middle frequency band, the uplink performance is negatively impacted due to higher path loss associated with higher frequencies, which degrade the users’ QoE in less optimal conditions. With the growing demand for high performance uplink throughput from novel applications such as Metaverse, Internet of Things (IoTs) and Smart City, 3GPP introduced High Power User Equipment (HPUE) on 5G TDD bands, allowing UEs to utilize more than 23 dBm of power for transmission to improve throughput, QoE, and reliability, especially at the cell edges. In this paper, the performance of HPUE is evaluated in the urban area on a commercial 5G network in terms of Uplink Throughput, Modulation Efficiency, Re-transmission Rate (ReTx Rate), and Power Consumption in both Standalone (SA) and Non-Standalone (NSA) modes. Through modem firmware modification, the performance is also compared across different power classes and antenna configurations.
\end{abstract}




%
\IEEEpeerreviewmaketitle

\vspace{-1mm}
\section{Introduction}
\begin{figure}[t!]
\centering\includesvg[width=0.98\linewidth,inkscapelatex=false]{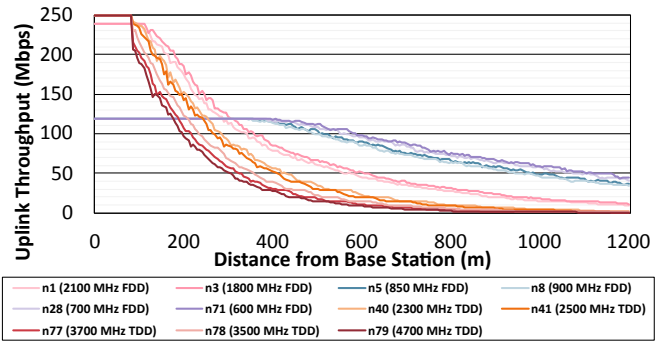}
\vspace{-1mm}
\caption{\textbf{Simulation}: Uplink Throughput (Mbps) vs Distance (m) by Frequency Bands. Dual-Tx for Frequency Bands \textgreater1 GHz and Power Class 3 were assumed.}
\label{fig:BandCompare}
\vspace{-8mm}
\end{figure}

 \setstretch{0.95}
The C-Band (3.5 GHz) and BRS (2.5 GHz) frequency bands are commonly allocated for Time Division Duplexing (TDD) 5G NR networks, which operate at a higher frequency compared to the bands commonly allocated for 4G LTE \cite{9052737}. This poses deployment challenges due to greater path loss associated with higher carrier frequencies \cite{8928031, 9135487}. While such issue can easily be mitigate in the downlink direction thanks to the higher transmission power budget of the base stations, typically in the range of 43 to 51 dBm in urban areas, as well as techniques like beamforming to concentrate the power in the direction of the User Equipment (UE) \cite{ericsson}, these techniques could not be performed effectively on the User Equipment (UE) side due to limited number of antenna with low gain. As a result, the uplink performance on middle-frequency bands commonly used for TDD 5G NR network degrades more rapidly compared to the FDD bands used in 4G Long Term Evolution (LTE) in less optimal conditions (as seen in Fig. \ref{fig:BandCompare}), causing poor users’ Quality of Experience (QoE) \cite{10570635}. With the new use cases such as Metaverse, Artificial Intelligence, Ultra High Definition (UHD) live streaming, and Massive Machine Type Communications (MTC), the demand for high-performance uplink on mobile networks surged significantly, especially in the urban areas, which is the target area for smart city implementation \cite{10373900,10574348}.

To address the uplink limitations of TDD NR frequency bands, 3GPP introduced the High-Power User Equipment (HPUE) feature for TDD NR bands \cite{3GPP_38-101-1}. At the time of this study, certain countries permitted HPUE operation in specific bands, including n41 (2.5 GHz), n77 (3.7 GHz), n78 (3.5 GHz), and n79 (4.9 GHz). HPUE is categorized into different power classes, such as Power Class 2 (PC2) and Power Class 1.5 (PC1.5), with maximum transmission power levels of 26 dBm and 29 dBm, respectively. By increasing uplink transmit power, HPUE enhances Quality of Experience (QoE) in both urban and cell-edge scenarios. In urban environments, higher transmission power increases the likelihood of utilizing higher Modulation and Coding Schemes (MCS), improving spectral efficiency and overall cell throughput. At the cell edge, higher power classes enhance communication reliability by reducing the Re-transmission Rate (ReTx Rate), thereby lowering latency and improving QoE for these users.

Since HPUE is expected to be a key enabler for reliable mobile services and massive communication in Smart City on middle-frequency TDD bands \cite{mediatek_2018}, real-world studies as well as further simulations are needed to understand its impact, benefits, and challenges. This paper analyzes the scalability of uplink performance across different power classes and antenna configurations, focusing on throughput gains, modulation scheme utilization, and ReTx rates in urban environments. Furthermore, the power consumption analysis are provided, based on measured UE power consumption across different common use cases and average Physical Uplink Shared Channel (PUSCH) transmit power obtained from field tests, to assess the impact of HPUE on perceived battery life. Finally, based on the results, recommendations are provided to assist Mobile Network Operators (MNO) with their network configurations and optimizations.\looseness=-1


\section{Experiment Environment}

\subsection{User Equipment (UE)}

The ASUS Smartphone for Snapdragon Insiders (EXP21), equipped with the Qualcomm Snapdragon X60 5G Modem-RF system \cite{qualcomm_x60}, was used as the user equipment (UE). The device was validated to ensure compatibility with the required frequency bands and Dual-Tx configuration, as Power Class 1.5 (PC1.5) operation is supported only in Dual-Tx mode \cite{3GPP_38-101-3}. The modem firmware was adjusted to override the network-defined \textit{p-max} value, allowing operation at the specified power class for each experiment. Moreover, this method also enabled switching between Single-Tx and Dual-Tx configurations as needed. As the \textit{q-RxLevMin} parameter was configured on 5G SA cells with the value being set to -96 dBm, the UE wouldn't be able to attach to the cell if the measured RSRP falls below the threshold, preventing the collection of data under the weaker signal conditions. Therefore, to combat this issue, modem firmware was also modified to ignore the \textit{q-RxLevMin} value, allowing the experiment UE to attach at any signal level. \looseness=-1

To ensure consistency, Uplink 256QAM was disabled due to the network’s use of Huawei’s \textit{UL\_256QAM\_ADAPT} feature, which dynamically assigns the 256QAM Modulation and Coding Scheme (MCS) Table (38.214 - Table 5.1.3.1-2) based on vendor's proprietary conditions. Finally, to confirm proper UE operation, the \textit{UECapabilityInformation} packet and maximum transmission power were monitored using professional drive test and modem log analysis tools, including \textit{Network Signal Guru (NSG)} and \textit{Qualcomm Commercial Analysis Toolkit (QCAT)}. \looseness=-1


\subsection{Network Environment}

\begin{table}[!tbp]
\setstretch{0.75}
\vspace{1.5mm}
\caption{Summary of AIS Network Frequency Bands}
\vspace{-1.5mm}
\centering
\label{tab-AISFreqBand}
\resizebox{8.5cm}{!}{\begin{tabular}{@{}lcccc@{}}
\toprule
Frequency Band                  &LTE B1&LTE B3& LTE B8   & 5G NR n41   \\\midrule
Duplex Mode                     &FDD&FDD& TDD       & TDD       \\
EARFCN/SSB-ARFCN & 350 & 1450 & 3650 & 506670/511950 \\
Downlink Carrier Frequency (MHz)  &2140-2150&1820-1840&940-950 & 2520-2580       \\
Uplink Carrier Frequency (MHz)    &1950-1960&1725-1745&895-905 & 2520-2580       \\
Channel Bandwidth (MHz)         &10&20& 10        & 60       \\
Sub Carrier Spacing (SCS)       &15 kHz&15 kHz& 15 kHz    & 30 kHz    \\
TDD Pattern Periodicity (ms) &N/A&N/A&N/A& 5\\
TDD Pattern Slots (DL/UL)     &N/A&N/A&N/A& 7/2\\
TDD Pattern Symbols (DL/UL)   &N/A&N/A&N/A& 6/6\\
Deployment Type & 4T4R & 4T4R & 4T4R & MaMIMO 64T64R\\
\bottomrule
\end{tabular}}
\vspace{-3mm}
\end{table}

\begin{figure}[!tbp]
    \centering\includesvg[width=0.95\linewidth,inkscapelatex=false]{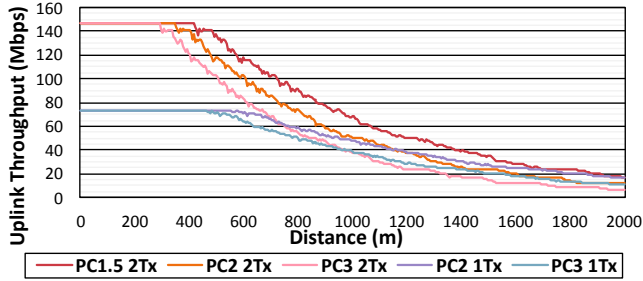}
    \vspace{-1mm}
    \caption{\textbf{Simulation}: 5G Uplink Throughput (Mbps) vs Distance (m) from Base Station (m) when using AIS configs.}
    \label{fig-DistanceThpt}
    \vspace{-5.5mm}
\end{figure}

\begin{table}[!tbp]
\setstretch{0.75}
\caption{\textbf{Field Test}: Summary of RF Conditions}
\vspace{-1.7mm}
\centering
\label{tab-UrbanCases}
\resizebox{8.5cm}{!}{\begin{tabular}{@{}ccccccccccc@{}}
\toprule
Power&MIMO&LTE&5G NR&Speed&LTE&LTE&LTE&5G NR&5G NR&5G NR \\
Class&&Band&Band&(km/h)&RSRP&SINR&PUCCH&SS-RSRP&SS-SINR&PUCCH \\
&&&&&(dBm)&(dB)&(dBm)&(dBm)&(dB)&(dBm)\\\midrule
\multicolumn{11}{c}{5G Standalone (SA) Experiments}\\\midrule
PC1.5&2Tx&N/A&n41&22.66&\multicolumn{3}{c}{\textit{Not Available (N/A)}}&-76.83&21.88&0.82 \\
PC2&2Tx&N/A&n41&19.11&\multicolumn{3}{c}{\textit{Not Available (N/A)}}&-76.15&21.77&0.11   \\
PC3&2Tx&N/A&n41&22.64&\multicolumn{3}{c}{\textit{Not Available (N/A)}}&-77.17&21.96&2.11   \\
PC2&1Tx&N/A&n41&17.63&\multicolumn{3}{c}{\textit{Not Available (N/A)}}&-74.60&22.13&-1.24  \\
PC3&1Tx&N/A&n41&17.94&\multicolumn{3}{c}{\textit{Not Available (N/A)}}&-75.02&21.61&0.43   \\\midrule
\multicolumn{11}{c}{5G Non-Standalone (NSA) Experiments}\\\midrule
PC2&NSA&B1&n41&18.15&-81.48&11.30&-16.72&-76.92&20.69&1.60  \\
PC3&NSA&B1&n41&17.76&-86.81&10.63&-9.96&-83.17&20.53&3.49   \\
PC2&NSA&B3&n41&19.57&-75.40&9.03&-13.44&-79.70&20.17&0.41   \\
PC3&NSA&B3&n41&21.12&-77.84&8.35&-10.88&-83.61&20.27&2.56   \\
PC2&NSA&B8&n41&18.16&-67.58&10.79&-21.25&-76.82&20.80&1.36  \\
PC3&NSA&B8&n41&22.32&-77.19&9.96&-14.12&-78.06&20.88&1.38   \\
\bottomrule
\end{tabular}}
\vspace{-4mm}
\end{table}

The field test was conducted in Bangkok, Thailand during June and October 2024 on the Advanced Info Service (AIS), which has the 5G services on the BRS frequency (2.5 GHz) or frequency band n41. At the time of the study, Power Class 2 (PC2) was supported and available in both SA and NSA mode. Approximately 95\% of AIS base stations were equipped with Huawei 64T64R Massive MIMO Active Antenna Units (AAUs), while the remaining base stations, mostly small cells, were deployed using Remote Radio Unit (RRU) + 4T4R passive antenna configuration. It should be noted that throughout the experiments, the UE was connected to base stations equipped with Massive MIMO AAUs. Moreover, commercial SIM cards and service plans were used in the study to emulate typical user conditions. For the 5G NSA experiments, three LTE anchor bands were utilized: Band 1 (2100 MHz), Band 3 (1800 MHz), and Band 8 (900 MHz). A detailed summary of the RF bands and TDD timing configurations used by the network in the study is provided in Table \ref{tab-AISFreqBand}.

\begin{figure}[!tbp]
    \centering
    \includegraphics[width=0.95\linewidth]{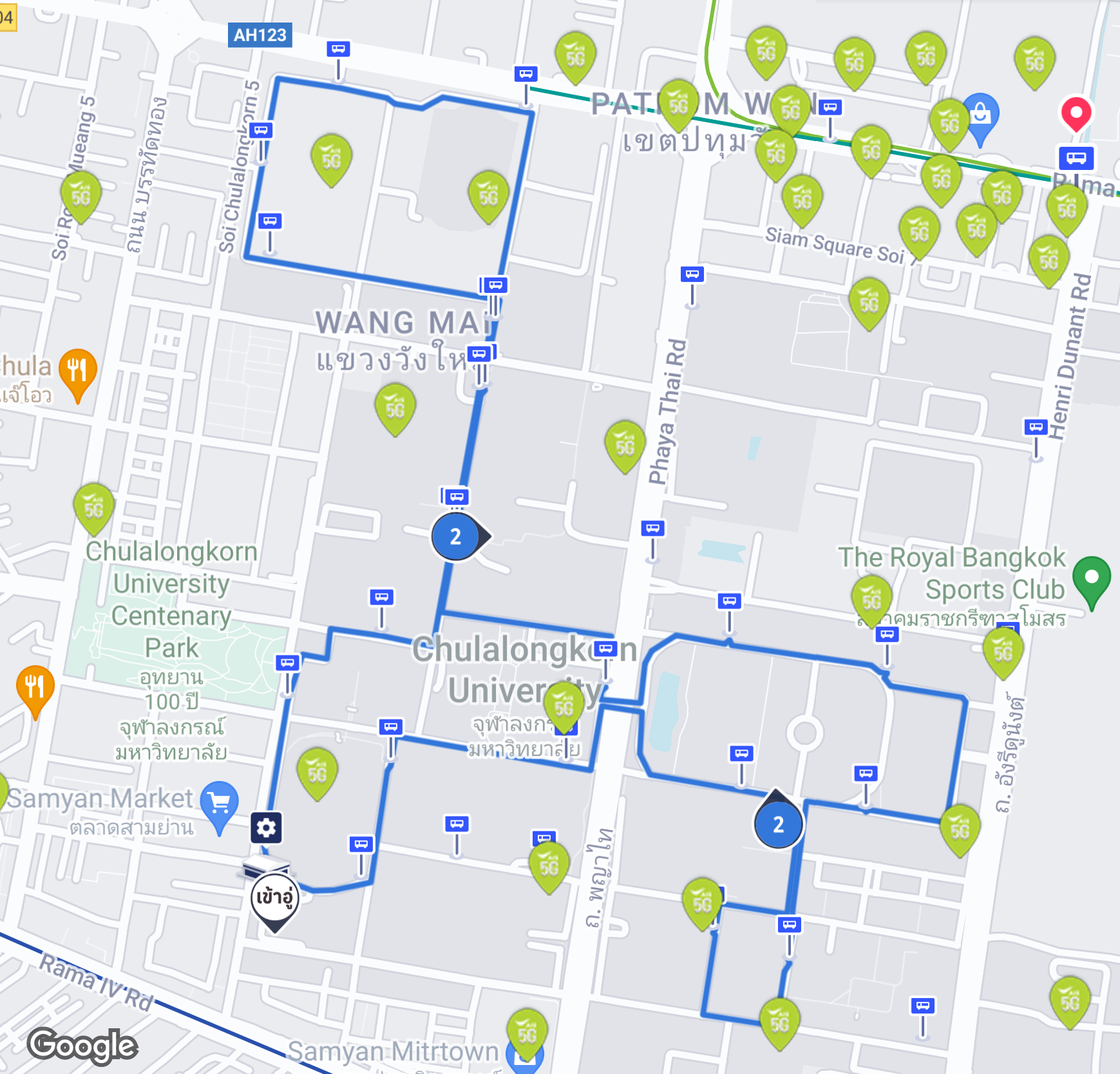}
    \vspace{-1mm}
    \caption{\textbf{Field Test}: Map of Pop Bus Route Two. Green pins represents gNodeB locations.}
    \label{fig-MapPopBus}
    \vspace{-6.5mm}
\end{figure}
\begin{figure}[!tbp]
    \centering
    \includegraphics[width=0.95\linewidth]{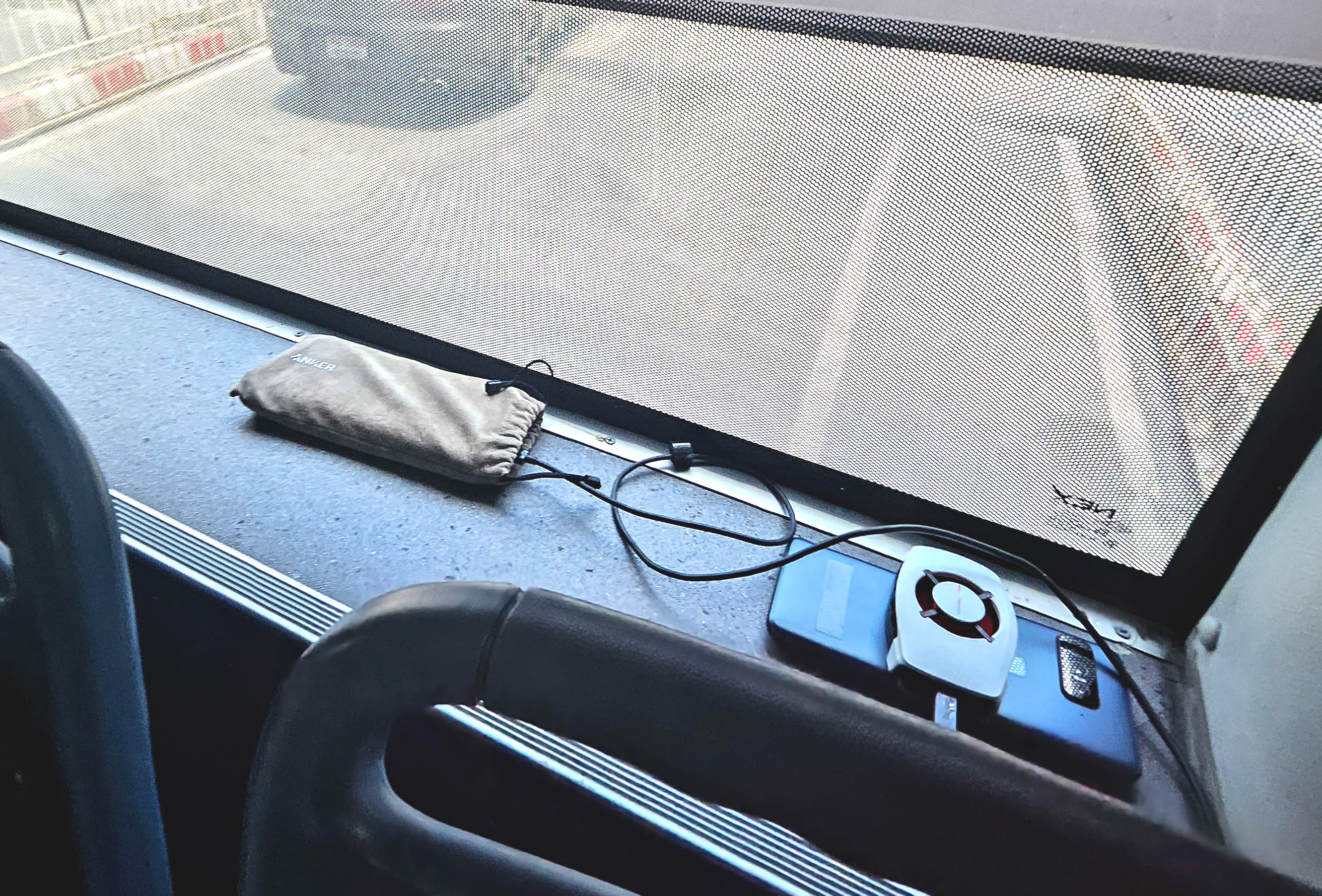}
    \vspace{-1mm}
    \caption{\textbf{Field Test}: UE Placement}
    \label{fig-PopBusBack}
    \vspace{-5mm}
\end{figure}

Due to the local laws and regulations, specifically the PC1.5 experiments, which exceeded the \textit{p-max} set by the network, the experimental area was restricted to \textit{Chulalongkorn University}. The university, located in the Pathum Wan district of central Bangkok, has a high base station density of 12.7 gNB per km\textsuperscript{2} (see Fig. \ref{fig-MapPopBus}). Experiments were conducted on the university's shuttle service, the \textit{"Pop Bus"}, specifically on route number two, which follows a 10 km loop with an average duration of approximately 25 minutes per lap. The bus maintained an average speed of 12 km/h. To ensure consistent network load across trials, experiments were conducted on weekdays between 9:00–12:00 and 13:00–15:00, thereby avoiding peak traffic periods such as lunch breaks and rush hours. The user equipment (UE) was equipped with an 18W thermoelectric cooler to mitigate potential performance degradation caused by thermal throttling. For each trial, the UE was positioned near the rear window of the bus (see Fig. \ref{fig-PopBusBack}). A total of 11 test scenarios were performed, comprising three 5G SA Dual-Tx, two 5G SA Single-Tx, and six 5G NSA configurations (see Table \ref{tab-UrbanCases}). During each test, NSG’s testing function was used to continuously transmitted data to AIS’s Ookla Speedtest server via the HTTP POST protocol, ensuring maximum radio link utilization. Given the variability in road traffic conditions and stop durations across trials, data points were averaged and re-sampled at eight-meter intervals before analysis.\looseness=-1

\section{Results and Analysis}
\subsection{Uplink Throughput and Modulation}

\begin{figure}[t!]
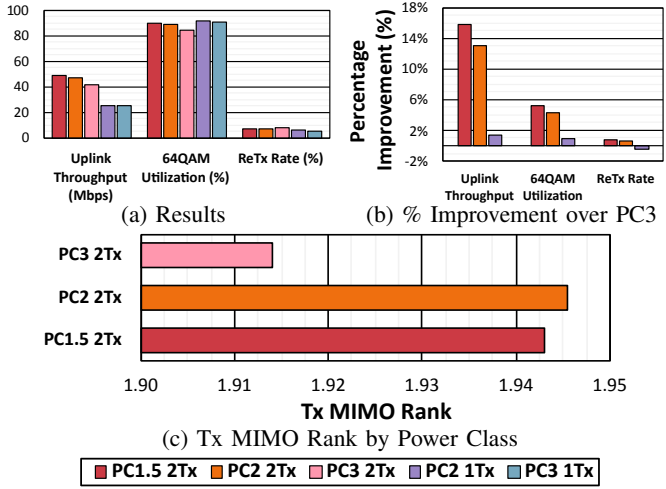

\centering
\begin{subfigure}[t]{0.50\linewidth}
\centering\includesvg[width=0.97\linewidth,inkscapelatex=false]{PopBusSummarySA.svg}
  \vspace{-2mm}
  \caption{Results}
  \label{fig-PopBusSummarySA}
\end{subfigure}\hfill%
\begin{subfigure}[t]{0.485\linewidth}
\centering\includesvg[width=0.97\linewidth,inkscapelatex=false]{PopBusImprovementSA.svg}
  \vspace{-2mm}
  \caption{\% Improvement over PC3}
  \label{fig-PopBusImprovementSA}
  \vspace{0.5mm}
\end{subfigure}\\
\begin{subfigure}[t]{0.9\linewidth}
\centering\includesvg[width=0.97\linewidth,inkscapelatex=false]{TxMIMORank.svg}
  \vspace{-1mm}
  \caption{Tx MIMO Rank by Power Class}
  \label{fig-TxMIMORank}
  \vspace{0.5mm}
\end{subfigure}
\\
\begin{subfigure}[t]{\linewidth}
\centering\includesvg[width=0.78\linewidth,inkscapelatex=false]{PowerClassDescriptionHorizontal.svg}
\end{subfigure}
\vspace{-5mm}
\caption{\textbf{Field Test}: 5G Standalone (SA) Results}
\vspace{-5.5mm}
\label{fig:PopBusSA}
\end{figure}

\begin{figure}[!tbp]
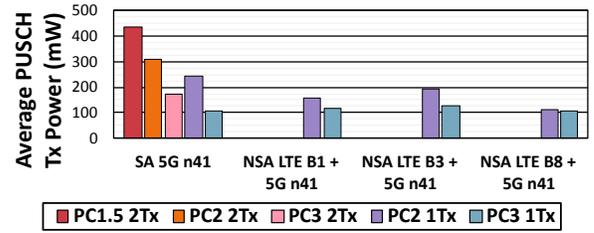

    \centering
    \includesvg[width=0.86\linewidth,inkscapelatex=false]{PowerUrban.svg}
    \vspace{-0.5mm}
    \begin{subfigure}[t]{\linewidth}
    \vspace{-2.5mm}
\centering\includesvg[width=0.78\linewidth,inkscapelatex=false]{PowerClassDescriptionHorizontal.svg}
\end{subfigure}
    \caption{\textbf{Field Test}: Average PUSCH Transmission Power. For 5G NSA scenarios, the reported power represents the combined transmission power of the SCG LTE carrier and the MCG 5G NR carrier. Data is presented in a linear scale to facilitate power consumption comparisons.}
    
    \label{fig-UrbanPwr}
    \vspace{-5.5mm}
\end{figure}

\begin{figure}[t!]
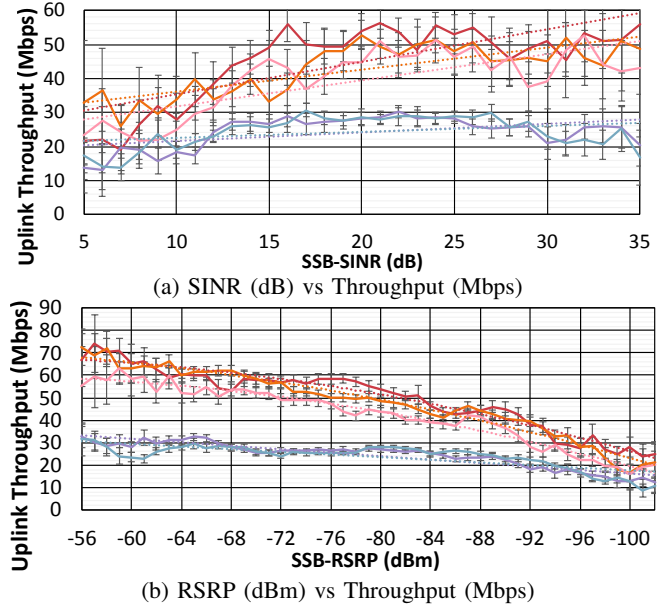

\begin{subfigure}{\linewidth}
\centering\includesvg[width=0.949\linewidth,inkscapelatex=false]{HPUE_SINR_v2.svg}
  \vspace{-1.6mm}
  \caption{SINR (dB) vs Throughput (Mbps)}
  \vspace{-0.2mm}
  \label{fig-thpt-SINR}
\end{subfigure}\\
\begin{subfigure}{\linewidth}
  \centering\includesvg[width=0.97\linewidth,inkscapelatex=false]{HPUE_RSRP_v2.svg}
  \vspace{-1mm}
  \caption{RSRP (dBm) vs Throughput (Mbps)}
  \label{fig-thpt-RSRP}
\end{subfigure}\\
\begin{subfigure}{\linewidth}
\centering
\vspace{0.4mm}
\includesvg[width=0.949\linewidth,inkscapelatex=false]{CurveDescription.svg}
\end{subfigure}
\vspace{-5mm}
\caption{\textbf{Field Test}: Throughput Behavior by Power Class. Error Bars represents 95\% Confidence Interval.}
\vspace{-7mm}
\label{fig:ThptCurvePopBus}
\end{figure}

One of the key benefits of using HPUE in dense urban areas is to improve both the spectral efficiency and users' throughput. Hence, this subsection analyzes the impact of HPUE in terms of uplink throughput and modulation across various UE configurations, for both 5G Standalone (SA) and Non-Standalone (NSA) deployments. Both of these metrics are critical indicators of spectral efficiency and users' QoE.

In SA mode, Single-Tx UEs showed minimal scaling with HPUE enabled. As shown in Fig. \ref{fig-PopBusSummarySA}, using PC2 in Single-Tx devices led to a barely perceptible 0.96\% increase in 64QAM utilization and a marginal 1.42\% improvement in uplink throughput over the baseline PC3 configuration. This limited gain occurs due to PC3 already providing enough power budget for Single-Tx UEs in this test scenario to sustain high-order modulations and operate near the upper limit of MCS index. The 64QAM utilization rate shows that Single-Tx UEs had already achieved 90.7\% 64QAM utilization at PC3, approaching the maximum spectral efficiency for MCS Table 1. Furthermore, the recorded PUSCH Tx Power also painted a similar picture, with the standard PC3 UE allocating an average transmission power of 20.23 dBm out of the allowed 23 dBm power budget (see Fig. \ref{fig-UrbanPwr}). Enabling PC2 increased the average PUSCH transmission power to 23.86 dBm, only slightly exceeding the PC3 power budget. This suggests that Single-Tx UEs seldom fully exploit the extra transmit power available at higher power classes, suggesting that PC3 is adequate in most cases for this particular test scenario. Therefore, increasing the power class provides minimal benefit and may result in unnecessary power consumption and interference. However, a slight benefit may be observed when using HPUE PC2 on a Single-Tx UE in combination with MCS Table 2. \looseness=-1


On the other hand, Dual-Tx UE results demonstrate the benefit of HPUE in urban environments more clearly. The results (see Fig. \ref{fig-PopBusSummarySA}) show that the PC3 Dual-Tx UE has a 6.35\% lower 64QAM utilization rate compared to Single-Tx PC3 UE, indicating an insufficient transmission power leading to a power bottleneck. This occurs due to the power budget being divided across two antennas, resulting in an effective power of 20 dBm per Tx chain. By enabling HPUE and using higher power classes like PC2 and PC1.5, the higher per-antenna power is restored to 23 dBm and 26 dBm, respectively, leading to measurable gains. Dual-Tx with PC2 shows a 4.30\% and 1.62\% increase in 64QAM and MIMO Rank 2 utilization rate, respectively, culminating in about 13.0\% uplink throughput uplift compared to the baseline Dual-Tx PC3 counterpart. With PC1.5, the throughput improvement increased further to 15.7\% over the baseline. \looseness=-1

While using the uplink RSRP and SINR would be a better metric for evaluating the uplink channel condition, such information is only obtainable on the base station side, which is not possible as it is considered confidential by the MNO. Therefore, the best two obtainable parameters—Downlink Reference Signals Received Power (RSRP) and Signal-to-Interference-plus-Noise Ratio (SINR) of the Synchronization Signal Block (SSB)—were used instead. Since the base stations in the area operate at a similar power level of around 47 dBm, RSRP provides a good representation of the amount of path loss at a given time, while SINR illustrates the channel condition at that time. Fig. \ref{fig:ThptCurvePopBus} shows the uplink throughput of all power classes across all antenna configurations by each level of measured downlink RSRP and SINR with a 95\% confidence interval. The result confirms the earlier findings, as it was found that the throughput curves of both PC2 and the baseline PC3 closely follow each other across all RSRP and SINR levels, demonstrating that the baseline PC3 UE is already sufficient for the urban environment and that using PC2 only leads to increased power consumption and potentially increased interference. However, the same could not be said for the Dual-Tx scenario, as the higher power class configurations perform better than the lower power classes most of the time across the measured signal levels, confirming that the additional permitted Tx power budget is exploited by Dual-Tx, effectively being translated into additional uplink throughput and improved spectral efficiency even in a high-base-station-density urban environment.

\begin{figure}[t!]
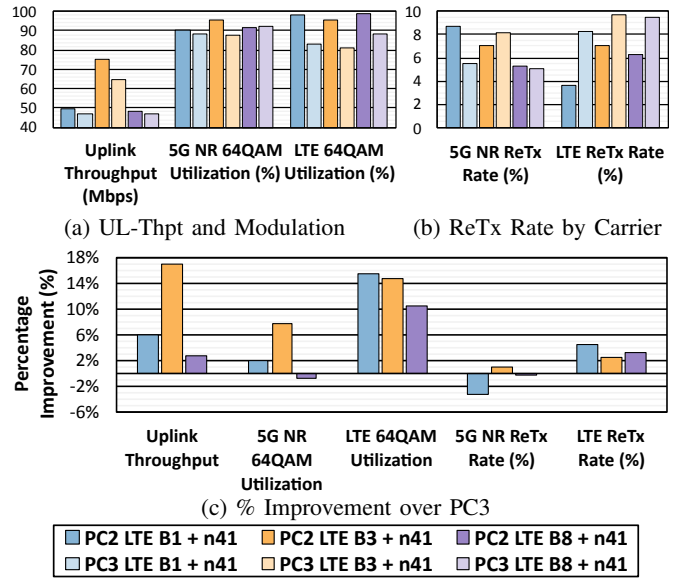

\begin{subfigure}{0.576\linewidth}
\centering\includesvg[width=\linewidth,inkscapelatex=false]{PopBusSummaryNSA-1.svg}
  \vspace{-4.8mm}
  \caption{UL-Thpt and Modulation}
  \vspace{-4mm}
  \label{fig-PopBusSummaryNSA-1}
  
\end{subfigure}
\begin{subfigure}{0.395\linewidth}
\centering\includesvg[width=\linewidth,inkscapelatex=false]{PopBusSummaryNSA-2.svg}
  \vspace{-2.0mm}
  \caption{ReTx Rate by Carrier}
  \vspace{-4mm}
  \label{fig-PopBusSummaryNSA-2}
\end{subfigure}
\\
\begin{subfigure}{\linewidth}
\vspace{5mm}
\centering\includesvg[width=0.98\linewidth,inkscapelatex=false]{PopBusImprovementNSA.svg}
  \vspace{-1.3mm}
  \caption{\% Improvement over PC3}
  \label{fig-PopBusImprovementNSA}
  \vspace{1mm}
\end{subfigure}\\
\begin{subfigure}[t]{\linewidth}
\vspace{-5mm}
\centering\includesvg[width=0.88\linewidth,inkscapelatex=false]{PowerClassDescriptionHorizontalNSA.svg}
\end{subfigure}
\vspace{-3.5mm}
\caption{\textbf{Field Test}: 5G Non-Standalone (NSA) Results}
\vspace{-6.5mm}
\label{fig:PopBusNSA}
\end{figure}

In NSA mode, a similar trend to that observed in Dual-Tx scenarios was identified. As shown in Fig. \ref{fig-PopBusSummaryNSA-1}, enabling PC2 led to a noticeable increase in throughput, particularly in 5G E-UTRAN New Radio Dual Connectivity (EN-DC) configurations that involve LTE carriers with larger channel bandwidths. The most significant gain, 16.9\%, was observed in the DC\_3A\_n41A configuration (LTE Band 3 + 5G Band n41), attributed to AIS's LTE Band 3 operating with a 20 MHz channel bandwidth. In contrast, configurations with narrower LTE channel bandwidths, such as DC\_1A\_n41A (LTE Band 1 + 5G Band n41) and DC\_8A\_n41A (LTE Band 8 + 5G Band n41), where the LTE carrier bandwidth is limited to 10 MHz, exhibited lower throughput gains of 6.13\% and 2.89\%, respectively. \looseness=-1

Furthermore, the utilization rate of 64QAM modulation on the LTE anchor carrier increased substantially when PC2 was enabled, ranging from 14.80\% to 15.46\% in configurations involving mid-frequency LTE carriers. However, in configurations utilizing low-frequency LTE carriers, such as DC\_8A\_n41A, a more limited gain of 10.44\% was observed. This outcome is attributed to the lower path loss associated with the lower carrier frequency of LTE Band 8 (900 MHz), which offers favorable propagation characteristics. Under the PC3 configuration, the LTE carrier’s 64-QAM utilization rate for the DC\_8A\_n41A configuration was already significantly higher (88.22\%) than that of configurations involving mid-frequency LTE carriers (80.96\% and 82.92\%). Enabling PC2 further increased the 64QAM utilization rate for the DC\_8A\_n41A configuration to 98.7\%, nearing the theoretical maximum spectral efficiency, thereby limiting additional benefits. \looseness=-1

Although channel bandwidth and frequency allocation vary across networks due to differences in local regulations, these findings are expected to be applicable to other networks worldwide with similar carrier frequencies and channel bandwidths. For instance, the results for the DC\_3A\_n41A configuration (LTE Band 3 – 1800 MHz) on AIS are likely comparable to those of the DC\_66A\_n41A configuration (LTE Band 66 – 1700 MHz) on T-Mobile US, as both LTE bands have similar carrier frequencies and operate with a 20 MHz channel bandwidth.\looseness=-1

\subsection{Re-Transmission Rate}

Re-transmission Rate (ReTx Rate) is a crucial metric reflecting link reliability and latency. Lowering the ReTx rate has always been one of the key goals in the field of telecommunication as it leads to higher connection reliability and improved users' QoE. \looseness=-1

In 5G SA mode, stepping up from PC3 to PC2 on Dual-Tx UEs results in a modest 0.58\% reduction in ReTx rate (see Fig. \ref{fig-PopBusImprovementSA}). This improvement suggests that the increased power budget from PC2 enhances link reliability for Dual-Tx UEs by improving the signal quality received by the gNB. With a stronger signal, the probability of successful initial decoding increases, leading to fewer initial transmission errors, which reduces the need for Hybrid Automatic Repeat Request (HARQ) re-transmission. This more efficient HARQ operation ultimately contributes to lower latency and increased link stability. However, the transition from PC2 to PC1.5 yields only a marginal 0.18\% further reduction in ReTx rate, indicating diminishing returns in reliability gains beyond PC2 in urban environments. This implies that while PC2 significantly improves reliability over PC3 for Dual-Tx UEs, the signal quality delivered by PC2 might already be approaching a threshold where further power increases from PC1.5 may only provide limited additional benefit in error probability and, consequently, ReTx rate under typical urban conditions. Furthermore, factors beyond signal power, such as residual interference or the inherent error correction limits of LDPC codes, might become more dominant in determining the ReTx rate at these higher power levels.

Conversely, for Single-Tx UEs, Fig. \ref{fig-PopBusImprovementSA} reveals that enabling PC2 and PC1.5 does not reduce the ReTx Rate. This observation aligns with the throughput results and further supports the findings that PC3 is often sufficient for Single-Tx UEs operating in urban environments both in terms of uplink throughput and reliability. The existing PC3 power level already provides adequate signal quality for Single-Tx UEs to achieve efficient HARQ operation and low ReTx rates under typical urban signal conditions. Hence, the extra power from PC2 does not translate into a measurable improvement in link reliability for Single-Tx UEs in these scenarios.

In 5G NSA mode, Results in Fig. \ref{fig-PopBusSummaryNSA-2} demonstrate that enabling PC2 primarily improves the ReTx rate of the LTE anchor carrier, with reductions ranging from 3.20\% to 4.63\%, while the 5G NR carrier displays negligible improvement in ReTx rate. This indicates the prioritization of the LTE anchor carrier's reliability within the NSA framework. This prioritization is logical and crucial as the LTE carrier in NSA mode operates as the Master Cell Group (MCG), carrying critical control and signaling information essential for overall connection stability. Improving the reliability of the MCG through HPUE ensures robust delivery of control signals, which is critical for maintaining a stable and reliable NSA connection. The negligible improvement in ReTx rate for the 5G NR carrier (Secondary Cell Group or SCG), which primarily handles user data, might suggest that the power split in NSA or other configuration factors limits the benefit of HPUE on the SCG, or that the NR link was already operating with sufficiently low ReTx rates even at baseline PC3.

\subsection{Power Consumption}


Enabling HPUE inevitably increases power consumption, which is a critical concern for battery-powered mobile devices such as smartphones and tablets, where battery life plays a crucial role. Fig. \ref{fig-UrbanPwr} presents the average power consumption for transmitting user data via the PUSCH across various power classes and antenna configurations. To assess the trade-off between increased power consumption and battery life, an analysis was conducted, as summarized in Table \ref{tab-battery}. For generalization, the battery capacity is assumed to be 16.7 Wh (4500 mAh @ 3.7 V), representing the average among commercially available 5G UEs. Three usage scenarios were considered: web browsing, video calling, and gaming, with estimated base power consumption (excluding RF power usage) of 1.5 W, 3.5 W, and 5.0 W, respectively. These estimations are derived from real-world measurements of multiple 5G UEs.

\begin{table}[!tbp]
\setstretch{0.75}
\vspace{1.5mm}
\caption{Estimated Battery Life on Various Use Cases by UE Power Class Configurations}
\vspace{-1.5mm}
\centering
\label{tab-battery}
\resizebox{8cm}{!}{\begin{tabular}{@{}lccc@{}}
\toprule
Uplink Configuration   &Web Browsing (h)&Video Calling (h)& Gaming (h)
\\\midrule
\multicolumn{4}{c}{5G Standalone (SA)}\\\midrule
\textbf{PC3 1Tx (Baseline)} &\textbf{10.40}&\textbf{4.63}&\textbf{3.27}\\
PC2 1Tx &9.58&4.46&3.19\\
\midrule
\textbf{PC3 2Tx (Baseline)}&\textbf{9.98}&\textbf{4.55}&\textbf{3.23}\\
PC2 2Tx&9.24&4.39&3.15\\
PC1.5&8.63&4.24&3.07\\
\midrule
\multicolumn{4}{c}{5G Non-Standalone (NSA)}\\\midrule
DC\_1A\_n41A \\\midrule
\textbf{PC3 (Baseline)} &\textbf{10.33}&\textbf{4.62}&\textbf{3.26}\\
PC2 &10.07&4.57&3.24\\
\midrule
DC\_3A\_n41A \\\midrule
\textbf{PC3 (Baseline)} &\textbf{10.27}&\textbf{4.61}&\textbf{3.26}\\
PC2 &9.86&4.52&3.22\\
\midrule
DC\_8A\_n41A \\\midrule
\textbf{PC3 (Baseline)} &\textbf{10.39}&\textbf{4.63}&\textbf{3.27}\\
PC2 &10.35&4.62&3.27\\
\bottomrule
\end{tabular}}
\vspace{-8mm}
\end{table}

In 5G SA mode, PC2 Single-Tx UEs, with an average transmit power of 23.86 dBm (242.96 mW), are estimated to have battery life durations of 9.58 hours, 4.46 hours, and 3.19 hours for web browsing, video calling, and gaming, respectively. This represents a significant reduction compared to PC3 Single-Tx UEs, which operate at a lower transmit power of 20.23 dBm (105.47 mW) and achieve extended battery life: 10.40 hours for web browsing, 4.63 hours for video calling, and 3.27 hours for gaming. Notably, PC2 Single-Tx UEs consume 2.5 times more power for RF transmission than their PC3 counterparts, resulting in a 7.88\% and 3.67\% reduction in battery life for web browsing and video calling, respectively. However, for gaming, where the System-on-Chip (SoC) is the dominant power consumer, the increased power consumption of the modem has a negligible impact on overall battery life. \looseness=-1


As expected, the Dual-Tx configuration exhibits higher power consumption. Dual-Tx PC3 UEs, with an average transmit power of 22.40 dBm (173.58 mW), consume 80\% more power than Single-Tx PC3 UEs for PUSCH transmission, leading to a corresponding reduction in battery life: 4.03\%, 1.73\%, and 1.22\% shorter durations for web browsing, video calling, and gaming, respectively. Dual-Tx PC2 UEs, operating at 24.89 dBm (308.32 mW), experience a further reduction in battery life, with 11.15\%, 5.18\%, and 3.67\% shorter durations for web browsing, video calling, and gaming, respectively, compared to Single-Tx PC3 UEs. While Dual-Tx PC2 UEs consume only 20\% more power than Single-Tx PC2 UEs for its RF operations, they provide significantly higher uplink throughput gains, suggesting a more favorable power consumption-performance trade-off. However, Dual-Tx PC1.5 UEs, with the highest average transmit power of 26.38 dBm (434.53 mW), offer only marginal throughput gains over Dual-Tx PC2 UEs while further decreasing battery life. This results in 17.02\%, 8.42\%, and 6.11\% reductions in available time for web browsing, video calling, and gaming, respectively, compared to Single-Tx PC3 UEs. These findings demonstrate diminishing returns in power efficiency for PC1.5 Dual-Tx configurations in urban scenarios. While the reduction in gaming time may be imperceptible to users due to the relatively small contribution of RF power consumption compared to other components, RF power usage make up a larger portion of total power consumption during day-to-day activities such as web browsing and video calling. A significant increase in RF power consumption is expected to noticeably impact perceived battery life and degrade the overall QoE of smartphone users. 

In 5G NSA mode, different uplink configurations involving PC2 exhibit varying power consumption and battery life impacts. Generally, configurations utilizing middle frequency LTE bands experience greater reductions in battery life when PC2 is enabled compared to their low frequency band counterparts. This is primarily due to the higher power allocation required to compensate for the increased path loss associated with higher-frequency bands. For instance, in the uplink configuration DC\_3A\_n41A (LTE Band 3 – 1800 MHz), enabling PC2 increases total PUSCH transmission power from 21.00 dBm (125.77 mW) to 22.86 dBm (193.18 mW), reducing web browsing time by 4.00\%, while only slightly affecting video calling and gaming durations. In contrast, in configurations utilizing low-frequency LTE bands, such as DC\_8A\_n41A (LTE Band 8 – 900 MHz), PC3 UEs already operate near maximum spectral efficiency. Consequently, enabling PC2 does not prompt the UE to allocate significantly more power to RF transmission. As a result, only a marginal increase in total PUSCH transmission power is observed, from 20.28 dBm (106.56 mW) to 20.55 dBm (113.40 mW), leading to a negligible impact on battery life.

\vspace{0.5mm}

\section{Conclusions and Future Work}

In this paper, the performance of High Power User Equipment (HPUE) was evaluated in urban environments on a commercial 5G New Radio (NR) network. While HPUE is primarily designed to enhance reliability in cell-edge scenarios, results show that it can also provide modest performance benefits in high cell-density urban environments at a cost of some impact on the battery life. HPUE enhances uplink throughput and connection reliability by enabling higher Modulation and Coding Schemes (MCS), increasing MIMO rank 2 utilization, while reducing retransmission rates (ReTx), which can improve latency and reliability. In the Single-Tx case, it was found that HPUE provides little to no benefit while introducing more battery drain. As a result, the results suggest that HPUE in urban environments is only suitable for UEs with Dual-Tx capability operating in SA mode as the performance improvement is more noticeable, justifying the additional power consumption. For 5G NSA mode, marginal performance improvements were observed due to the power budget split, but instead of dividing the power budget for each MIMO Tx chain, it is being divided across LTE MCG and NR SCG. Overall, PC3 is sufficient for Single-Tx UEs operating in SA mode. However, for 5G SA Dual-Tx and 5G NSA modes, in which the power budget is divided between two Tx chains or carriers, respectively, are more likely to benefit from HPUE. Network operators can further optimize performance and mitigate inter-cell interference by tuning the \textit{p-max} parameter at the gNB to manage HPUE-enabled transmissions more effectively. \looseness=-1

However, it is well known in the previous literature that using HPUE will result in an increase in inter-cell interference (ICI), especially in high-density deployment scenarios with small Inter-Site Distance (ISD), which can degrade the uplink performance of UEs in other cell groups \cite{3GPP_36-886, 3GPP_38-861}. Therefore, further research will focus on understanding the impact of uplink ICI by utilizing network simulators. After that, methods to mitigate the negative impacts arising from HPUE should be proposed through network planning and other optimization strategies. For example, techniques such as Fractional Frequency Reuse (FFR) can be utilized to reduce interference at the cell's edge \cite{articleHetNet}, where maximum negative impact caused by HPUE was observed. \looseness=-1

\section*{Acknowledgement}

This paper is supported by the Ministry of Internal Affairs and Communications (MIC) Project for Efficient Frequency Utilization Toward Wireless IP Multicasting. Additionally, the authors would like to express their gratitude to \textbf{PEI Xiaohong} of \textit{Qtrun Technologies} for providing \textit{Network Signal Guru (NSG)} and \textit{AirScreen}, the cellular network drive test software used for result collection and analysis in this research.






%

\setstretch{0.9}
\renewcommand{\IEEEbibitemsep}{0pt plus 0.5pt}
\makeatletter
\IEEEtriggercmd{\looseness=-1}
\makeatother
\IEEEtriggeratref{1}
\Urlmuskip=0mu plus 1mu\relax

\bibliographystyle{IEEEtran}
\vspace{-0.5mm}
\bibliography{b_reference}

\end{document}